\documentclass[11pt]{article}
\usepackage{amsmath,amssymb,fullpage}
\usepackage{algorithm, pseudocode}

\def\K{\ensuremath{\mathbb{K}}}

\def\A{\ensuremath{\mathbb{A}}}
\def\M{\ensuremath{\mathsf{M}}}

\begin{document}

\title{A fast algorithm for solving linearly recurrent sequences}
\author{Seung Gyu Hyun, Stephen Melczer, Catherine St-Pierre}
\date{}

\maketitle

\begin{abstract}
We present an algorithm which computes the $D^{th}$ term of a sequence satisfying a linear recurrence relation of order $d$ over a field $\K$ in $O( \M(\bar d)\log(D) + \M(d)\log(d))$ operations in $\K$, where $\bar d \leq d$ is the degree of the squarefree part of the annihilating polynomial of the recurrence and $\M$ is the cost of polynomial multiplication in $\K$. This is a refinement of the previously optimal result of $O( \M(d)\log(D) )$ operations, due to Fiduccia.
\end{abstract}

\paragraph{Overview.}
Consider a sequence $(a_i)_{i\ge 0}$ with entries in a field $\K$ that is generated by the recurrence
\begin{equation}\label{eq:rec}
a_{i+d} = \sum_{j=0}^{d-1} c_{j} a_{i+j}  
\end{equation}  
for all $i \ge 0$, where $c_0 \neq 0$.  Given initial conditions $a_j$, for $0 \le j \le d-1$, along with the annihilating polynomial $P = x^d - \sum_{j=0}^{d-1} c_{j} x^j$, we are interested in the complexity of computing one term $a_D$ of the sequence for some index $D \gg 0$.

A naive solution entails computing all terms $a_0,\dots,a_D$, but one can do much better. It has been known since at least Fiduccia's work~\cite{Fiduccia85} that computing $a_D$ can be reduced to multiplication modulo $P$. Explicitly, define $\A=\K[x]/P$, together with the $\K$-linear form $\ell: \A \to\K$ given by $\ell(x^i) = a_i$ for $i=0,\dots,d-1$. Then, since the residue class of $x$ in $\A$ is a root of $P$, its powers $(x^i)$ in $\A$ satisfy~\eqref{eq:rec}, and so do the values $(\ell(x^i))_{i \ge 0}$; this implies that $\ell(x^i)=a_i$ holds for all values of $i \ge 0$. In other words, to compute $a_D$ it is enough to compute $R=x^D \bmod P$ as $R = r_0 + \cdots + r_{d-1} x^{d-1}$, since then $a_D = r_0 a_0 + \cdots + r_{d-1} a_{d-1}$.

Letting $\M$ denote a function such that polynomials of degree $n$ can be multiplied in $\M(n)$ operations (under the assumptions of~\cite[Chapter~8]{GaGe13}), computing $R$ costs $O(\M(d)\log( D))$ operations in $\K$; we can then deduce $a_D$ in $O(d)$ steps.

\paragraph{A new algorithm.}
In this note, we present an improvement over this previous result that is useful when $P$ has multiple factors of high multiplicities.

First, we reduce to the case where $P$ has the form $P=Q^m$, for squarefree $Q$. To that end, let $P = \prod_i Q_i^{m_i}$ be the squarefree factorization of $P$, with pairwise distinct $m_1,\dots, m_s$ and $Q_i$ squarefree of degree $f_i$ for all $i$; note $d = \sum_i f_i m_i$.  We can compute $x^D \bmod P$ by applying the Chinese Remainder Theorem to the quantities $x^D \bmod Q_i^{m_i}$, giving the following algorithm.

\begin{algorithm}[H]
  \caption{Computing one element in a linear recurrent sequence}
  \textbf{Input:} 
  \begin{itemize}
  \item $P$: characteristic polynomial of the sequence
  \item $v=[a_0,\dots,a_{d-1}]$: vector of initial conditions
  \item $D$: index
  \end{itemize}
  \textbf{Output:} $D^{th}$ element of the sequence $(a_i)$ as in~\eqref{eq:rec}
  \begin{enumerate}
  \item compute the squarefree factorization of $P$ as $P = \prod_i Q_i^{m_i}$
  \item for $i=1,\dots,n$, compute $C_i = x^D \bmod Q_i^{m_i}$
  \item compute $R = x^D \bmod P$ by CRT as $R=r_0 + \cdots + r_{d-1} x^{d-1}$
  \item return $a_D =r_0 a_0 + \cdots + r_{d-1} a_{d-1}$.
  \end{enumerate}
\end{algorithm}

To compute the $C_i$'s efficiently, we will use bivariate computations. Indeed, for $i=1,\dots,s$ define $\A_i = \K[X]/Q_i^{m_i}$; then there exists a $\K$-algebra isomorphism 
$$ \pi_i: \A_i = \K[X]/Q_i^{m_i} \to \K[y,x]/\langle Q_i(y),(x-y)^{m_i} \rangle. $$ 
Following van der Hoeven and Lecerf~\cite{HoLe17}, we will call $\pi_i$ the operation of {\em untangling} (this is a conversion from a univariate representation to a bivariate one) and its inverse {\em tangling}. (To be precise, van der Hoeven and Lecerf consider a mapping to $\K[y,x]/\langle Q_i(y),x^{m_i} \rangle$, which is isomorphic to $\K[y,x]/\langle Q_i(y),(x-y)^{m_i}\rangle$ through the shift $x \mapsto x+y$).

van der Hoeven and Lecerf prove that for a given index $i$, untangling can be done in $O(\M(f_i m_i) \log(m_i))$ operations in $\K$; note that input and output sizes are $f_i m_i$ in this case. They also give an algorithm for tangling of cost $O(\M(f_i m_i)\log^2(m_i) + \M(f_i) \log(f_i))$; we give below a Las Vegas algorithm of cost $O(\M(f_i m_i) \log(f_i m_i))$ for this task. Taking the existence of such algorithms for granted, we write
$$ C_i = x^D \bmod Q_i^{m_i} = \pi_i^{-1}( \delta_i ),\quad\text{with}\quad \delta_i=x^D \bmod \langle Q_i(y),(x-y)^{m_i}\rangle. $$
The following allows us to compute $\delta_i$ efficiently. Define coefficients $e_0,\dots,e_{m_i-1}$ by
$$x^D \bmod (x-1)^{m_i} = e_0 + e_1x + \cdots + e_{m_i - 1} x^{m_i -1},$$
and define $S_i(x) = (yx)^D \bmod (x-1)^{m_i} \in \K[y][x]$, where $y$ is seen as an element of $\K[y]/Q_i(y)$. Then
$$\delta_i = x^D \bmod (x-y)^{m_i} = S_i\left(\frac{x}{y}\right),$$ 
where we note $y$ is invertible in $\K[y]/Q_i(y)$ as $c_0 \neq 0$. Now, 
$ S_i = y^D x^D \bmod (x-1)^{m_i}= y^D (e_0 + e_1x + \cdots + e_{m_i - 1} x^{m_i -1}),$ 
so that 
$$\delta_i = y^De_0 + y^{D-1}e_1x + \cdots + y^{D-(m_i - 1)}e_{m_i - 1} x^{m_i -1}.$$

In this algorithm, we first need to compute coefficients $e_0,\dots,e_{m_i-1}$; assuming that $2,\dots,m_i-1$ are units in $\K$, they can be obtained in $O(\log(D) + \M(m_i))$ operations in $\K$. The powers of $y$ we need are computed modulo $Q_i$, in time $O( \M(f_i)\log (D) + m_i \M(f_i))$. Altogether, we obtain $\delta_i$ using $O( \M(f_i)\log (D) + \M(f_i m_i) ))$ operations in $\K$. As said above, we can deduce $C_i$ from $\delta_i$ in Las Vegas time $O(\M(f_im_i) \log( f_im_i)).$ Taking all $i$'s into account and using the super-linearity of $\M$, the total time to compute $C_1,\dots,C_s$ is thus
$$O( \M(\bar d)\log(D) + \M(d)\log(d)),$$
where $\bar d =\sum_i f_i \le d$ is the degree of the squarefree part of $P$.

Computing the squarefree factorization of $P$ and Chinese remaindering both cost $O(\M(d)\log d)$ \cite[Corollary~10.23]{GaGe13}, so the overall runtime is $O( \M(\bar d)\log(D) + \M(d)\log(d))$. This is to be compared with the cost $O( \M(d)\log(D) )$ of Fiduccia's algorithm.

\paragraph{Tangling and untangling.}
We conclude by sketching our new algorithm for tangling. van der Hoeven and Lecerf reduce the tangling operation to untangling by means of a divide-and-conquer process; we propose a direct reduction that uses transposition, inspired by an algorithm from~\cite{Shoup94} that applies in univariate situations.

In what follows we use the same notation as in the previous paragraphs, but we drop the subscript $i$ for clarity. In particular, we write $f=\deg(Q)$ and $n=fm$ for the degree of $Q^m$, that is, the input and output size. Given $\delta$ in $\K[y,x]/\langle Q(y), (x-y)^m\rangle$,  we want to find $C=c_0 + \cdots + c_{n-1} x^{n-1}$ such that $\pi(C)=\delta$; this simply means that
$$ C \bmod \langle Q(y), (x-y)^m\rangle = \delta.$$
Choose a random linear form $\lambda:\K[y,x]/\langle Q(y), (x-y)^m\rangle \to\K$. For $j \ge 0$, multiply the former equality by $x^j$ and apply $\lambda$; this gives
$$c_0 \lambda(x^j) + \cdots + c_{n-1} \lambda(x^{j+n-1}) = \lambda (x^j \delta).$$
Taking $j=0,\dots,n-1$, we can collect these equalities in a linear system $H A = L$, with
\begin{align*} \label{matrix_prod}
H=      \begin{bmatrix}
          \lambda(1)  &  \hdots      & \lambda(x^{n-1}) \\
          \vdots & \ddots & \vdots \\
          \lambda(x^{n-1})&\hdots  & \lambda(x^{2n-2})\\
        \end{bmatrix}
\qquad
A=      \begin{bmatrix}
          c_0 \\ \vdots \\ c_{n-1}\\
        \end{bmatrix}
\qquad
L=      \begin{bmatrix}
          \lambda(\delta)\\ \vdots \\ \lambda(x^{n-1}\delta)\\
        \end{bmatrix}.
\end{align*}
Once $H$ and $L$ are known, we can recover coefficients $A$ in $O(\M(n)\log(n))$ operations in $\K$, since the linear system is Hankel (for a generic choice of $\lambda$, matrix $H$ has full rank $n$). Hence, the main question is the efficient computation of the entries of matrices $H$ and $L$. Both are instances of the same problem: given a $\K$-linear form $\lambda$ over $\K[y,x]/\langle Q(y), (x-y)^m\rangle$, and $\beta$ in  $\K[y,x]/\langle Q(y), (x-y)^m\rangle$, compute the values $\lambda(\beta),\dots,\lambda(x^{n-1}\beta)$.

As already recognized by Shoup in the univariate case, this question is the transpose of the untangling map $\pi^{-1}$. As a result, using the so-called {\em transposition principle}~\cite{Shoup94,Kaltofen00,BoLeSc03}, we can deduce an algorithm of cost $O(\M(n)\log(n)) = O(\M(f m) \log(f m))$ for tangling, by transposition of van der Hoeven and Lecerf's untangling algorithm.

\bibliographystyle{plain} 
\bibliography{abstract}

\begin{thebibliography}{1}

\bibitem{BoLeSc03}
A.~Bostan, G.~Lecerf, and {\'E}.~Schost.
\newblock Tellegen's principle into practice.
\newblock In {\em ISSAC'03}, pages 37--44. ACM, 2003.

\bibitem{Fiduccia85}
C.~M. Fiduccia.
\newblock An efficient formula for linear recurrences.
\newblock {\em SIAM Journal on Computing}, 14(1):106--112, 1985.

\bibitem{Kaltofen00}
E.~Kaltofen.
\newblock Challenges of symbolic computation: my favorite open problems.
\newblock {\em J. Symb. Comp.}, 29(6):891--919, 2000.

\bibitem{Shoup94}
V.~Shoup.
\newblock Fast construction of irreducible polynomials over finite fields.
\newblock {\em Journal of Symbolic Computation}, 17(5):371--391, 1994.

\bibitem{HoLe17}
J.~van~der Hoeven and G.~Lecerf.
\newblock Composition modulo powers of polynomials.
\newblock In {\em ISSAC '17}, pages 445--452. ACM, 2017.

\bibitem{GaGe13}
J.~von~zur Gathen and J.~Gerhard.
\newblock {\em Modern Computer Algebra}.
\newblock Cambridge University Press, Cambridge, third edition, 2013.

\end{thebibliography}

\end{document}